%% file: rate_adaptation.tex
\documentclass[12pt]{article}

\usepackage{amsthm}

\newtheorem{proposition}{Proposition}
\newtheorem{lemma}{Lemma}
\newtheorem{example}{Example}
\usepackage{graphicx}
\usepackage{mathrsfs}
\usepackage{amsmath}
\usepackage{amsfonts}
\usepackage{amssymb}
\usepackage{multirow}

\renewcommand{\P}{\ensuremath{\mathbb{P}}}

\newcommand{\N}{\ensuremath{\mathbb{N}}}

\newcommand{\comment}[1]{}

\begin{document}
\title{Social Attention and The Provider's Dilemma}
\author{Christina Aperjis and Bernardo A. Huberman \\ Social Computing Lab, HP Labs}
\date{}
\maketitle

\thispagestyle{empty}

\section*{Abstract}

While attracting attention is one of the prime goals of content providers, the conversion of that attention into revenue is by no means obvious. Given that most users expect to consume web content for free, a provider with an established audience faces a dilemma. Since the introduction of advertisements or subscription fees will be construed by users as an inconvenience which may lead them to stop using the site, what should the provider do in order to maximize revenues? We address this question through the lens of adaptation theory, which states that even though a change affects a person's utility initially, as time goes on people tend to adapt and become less aware of past changes.  We establish that if the likelihood of continuing to attend to the provider after an increase in inconvenience is log-concave in the magnitude of the increase, then the provider faces a tradeoff between achieving a higher revenue per user sooner and maximizing the number of users in the long term.  On the other hand, if the likelihood of continuing to attend to the provider after an increase in inconvenience is log-convex, then it is always optimal for the provider to perform the increase in one step. 
\newpage

\pagenumbering{arabic}

\section{Introduction}

While the explosion of content offered by the web constitutes a bonanza for consumers, the same cannot be said of the content providers. From the early days of the web, keen competition for the attention of users~\cite{falkinger:attention, wu, huberman:attention} dictated that  providers offer both access and consumption for free. As a result, most users expect not only to easily access any kind of content, but to consume it at no cost; an expectation that is embodied by the ``information wants to be free'' manifesto~\cite{brand:free_info}.

The goal of content providers is to turn attention to their sites into revenues that will at least offset their costs.
However, even providers with established audiences often struggle to convert the attention they receive into profits (e.g.,~\cite{LondonTimes}).
There are many ways of converting attention to revenue;  charging subscription rates and presenting adverts are typical examples.  Mixed strategies, where subscription fees and advertising are combined, have also been considered~\cite{baye:ad, prasad:advertising, kumar:control}.  But all these strategies carry a price, for while some users perceive the associated costs as  an inconvenience to be tolerated  in exchange for the value obtained,  others see them as a nuisance that makes them leave the site.  This issue has been especially acute with the advent of increasingly intrusive ``rich media'' online advertising formats~\cite{godes:MS}.

Given that a provider can increase his revenue by imposing some inconvenience to users while risking losing some of the attention paid to his content, how steeply and for how long should he increase this inconvenience in order to maximize revenue?
We address this question through the lens of adaptation theory, which states that even though a change affects a person's happiness in the short term, in the course of time people tend to adapt and become less aware of past changes~\cite{frederick, frey:hapiness}. Furthermore, as a number of empirical studies show,  gradual changes and spikes in utility have rather different effects on adaptation levels: whereas sudden changes are noticed and evaluated, a very slow gradual change will drag the adaptation level along with it and at times may not even be detected~\cite{kahneman:comp}.

We treat the dynamics of the adaptation process in settings where multiple changes in the utility to the user occur over time.  Intuitively, if the rate at which changes are introduced is much smaller than the rate at which people adapt to changes, then users will adapt over time and continue to give their attention to the site.  On the other hand, if the rate at which inconveniences are introduced is much higher than the rate at which people adapt, then users will leave the site. The interesting and challenging regime is the one where the two rates have the same order of magnitude, and as a result, some people are able to adapt while some cannot. In this case the provider may face a tradeoff between achieving a higher revenue per user sooner and maximizing the number of users in the long term.


Our analysis is based on the probability that a user continues to attend to the provider after an increase in inconvenience as a function of the magnitude of the increase; a function that can be measured in real settings (e.g., with A/B testing).
The shape of this function characterizes the strategy that maximizes the provider's revenue.
We find that if the likelihood of continuing to attend to the provider after an increase in inconvenience is log-convex (i.e., the logarithm of the function is convex), then it is always optimal for the provider to perform the increase in one step.

On the other hand, if the likelihood of continuing to attend to the provider after an increase in inconvenience is log-concave in the magnitude of the increase, then the provider faces a tradeoff between achieving a higher revenue per user sooner and maximizing the number of users in the long term.  Moreover, in the case of a fixed target level of inconvenience, the long-term number of users is a decreasing function of the average rate of increase in inconvenience.
We provide an algorithm for solving the revenue maximization problem of the provider.

The paper is structured as follows.  In Section \ref{sec:related} we review literature on maximizing the revenue of a content provider.  Section \ref{sec:model} introduces the model.
Sections \ref{sec:simple}, \ref{sec:discontinuity}, \ref{sec:time}, and \ref{sec:lasting} assume a fixed target increase in inconvenience and study monotonicity properties of the fraction of users that stay after the whole amount of inconvenience is introduced.  Section \ref{sec:simple} considers a setting where users completely adapt to changes and the time to adapt is independent of the magnitude of the change; two assumptions that we relax in the following sections.  Section \ref{sec:discontinuity} studies adaptation in the presence of discontinuous behavior.  Section \ref{sec:time} assumes that the time to adapt depends on the magnitude of the change.  Section \ref{sec:lasting} considers lasting effects, that is, settings where users do not completely adapt to changes.
Then, in Section \ref{sec:opt} we consider the provider's revenue optimization, where both the inconvenience increase per period and the total increase in inconvenience are chosen optimally.
Section \ref{sec:conclusion} concludes.  All the proofs are provided in the Appendix.

\section{Related Literature}
\label{sec:related}

Advertisements and subscription fees are the main sources of revenue for most content providers.  A number of studies have considered their optimal levels both separately and jointly.
Dewan et al. use optimal control to balance the amount of advertising and content on a web page~\cite{dewan:HICSS}.
Kumar and Sethi extend the optimal control model to also include subscription fees~\cite{kumar:control}.
Prasad et al. suggest that websites can increase their revenues by offering a menu of contracts, where high subscription fees are associated with a small number of ads and vice versa~\cite{prasad:advertising}.  Baye and Morgan present a model that explains why traditional and modern mass media --- such as television, newspapers, magazines, and Internet sites --- typically derive the bulk of their revenues from advertisements rather than subscriptions~\cite{baye:ad}.
Godes et al. explore the implications of two-sided competition for the sale of content to consumers and the attraction of advertisers on the actions and source of profits of media firms~\cite{godes:MS}.

Even though some of the aforementioned papers consider the effect of inconvenience on the number of users (e.g., \cite{godes:MS, kumar:control}), they do not consider the effects of adaptation.  To the best of our knowledge, this is the first paper to consider adaptation in the context of a content provider's revenue.  Adaptation theory allows us to consider how users react over time to an introduced inconvenience.
In this paper, we consider a general framework that applies to any type of inconvenience to the user that generates revenue for the provider.  Subscription fees and advertisements are typical examples.

\section{The Model}
\label{sec:model}

In this section we discuss our modeling assumptions with respect to the users and the providers.

\subsection{Users}
\label{sec:users}

Throughout the paper, we denote by $p(x)$ the probability that a user continues using the site after the inconvenience level (e.g., advertisement level, subscription cost) is increased by $x$. This probability captures heterogeneity in the user population: some users may be more likely to stay than others, and a user chosen at random stays with probability $p(x)$.

We assume that $p(0) = 1$, i.e., all users continue to use the site if there is no change.  Moreover, $p(x)$ is a decreasing function of $x$: the larger the increase in inconvenience, the smaller the probability of continuing to use the site.

We next describe a utility model that gives rise to such a probability $p(x)$.  Nevertheless, many of our results are stated in terms of $p(x)$ and hold even if this utility model does not apply.

\subsubsection{Additive Random Utility Model}
\label{sec:ARUM}

We consider the user's {\em experienced utility}, that is, the hedonic experience associated with the use of a website (see~\cite{anomalies}).
Let $u_0$ be the user's current level of utility from using the website, and assume that he gets zero utility from not using a website.  We further assume that $u_0 > 0$, so that the user is initially better off using the site.

We assume that if the inconvenience that a user experiences at the website increases by a strictly positive amount $x$, then the user incurs a cost of $c(x) + Y$, where $c(x)$ is an increasing function of $x$ and $Y$ is a random term drawn from a distribution whose cumulative distribution function is given by $F$.  Thus, the user continues to use the website after the undesired change with probability
\[p(x) = \P[u_0 - (c(x) + Y) > 0] = \P[Y < u_0 - c(x)] = F(u_0 - c(x)).\]
This Additive Random Utility Model (ARUM)\footnote{We note that ARUMs are often used in economics, see e.g.,~\cite{microeconometrics}.} and the assumption that $p(0) = 1$ imply that
\begin{equation*}
p(x)=
\begin{cases} F(u_0 - c(x)) & \text{if $x>0$,}
\\
1 &\text{if $x = 0$.}
\end{cases}
\end{equation*}

We note that a $p(x)$ that arises from this ARUM is decreasing in $x$, because $c(x)$ is an increasing function of $x$ and thus $F(u_0 - c(x))$ is a decreasing function of $x$.

We also point out that for any decreasing function $p(x)$ on $[0, \infty)$ such that $p(0) = 1$ and $p(x) \geq 0$, one can construct a utility model that generates it.  For example, this can be achieved by setting $u_0 = 1$, $c(x) = x$, and $F(y) = p(1-y)$ for $y \leq 1$.

\subsubsection{Adaptation}

According to adaptation theory, even though a change initially affects a person's happiness, as time goes on people tend to adapt and become less aware of past changes.  In the context of our theory, an increase in inconvenience by an amount $x$ initially decreases a user's utility by $c(x) + Y$; and we assume that as time goes by and if no additional inconvenience is experienced, the user's experienced utility gradually increases.  The user's utility may either increase up to $u_0$ --- complete adaptation --- or up to some smaller value $u_0$, which signals the existence of lasting effects.

In terms of the probability $p(x)$, if there is complete adaptation and a sufficient amount of time has elapsed since the last increase in inconvenience, then a current user will stay with probability $p(x)$ if inconvenience is increased by $x$.  The probability of staying will be smaller in the case of incomplete adaptation.  The latter is modeled in Section \ref{sec:lasting}.

\subsection{Provider}
\label{sec:providers}

In the first part of the paper (Sections \ref{sec:simple}, \ref{sec:discontinuity}, \ref{sec:time}, and \ref{sec:lasting}) we assume that the provider wishes to increase the inconvenience that users experience by some fixed amount $A$, and study monotonicity properties of the fraction of users that stay after the total increase $A$ is introduced.
There are of course many ways whereby the target inconvenience level can be reached.   For example, it may be reached through a single increase of $A$, through two increases of $A/2$, or through ten increases of $A/10$.

We note that $A$ is the increase in inconvenience, which in general may be different from the target inconvenience level.  Thus, if users currently do not experience any inconvenience, then $A$ is the target inconvenience level.  Otherwise, if users are already experiencing some inconvenience (say $A_0$), then the target inconvenience level is $A_0 + A$.  

In Section \ref{sec:opt}, we consider the provider's revenue optimization, where both $A$ and the number of increases through which it is introduced are chosen optimally.
The provider wishes to maximize his revenue, which at any given point in time is an increasing function of both the number of users and the current inconvenience level.
Furthermore,  we assume that the providers discounts future payments in that he prefers to get revenue sooner than later.

Key notation introduced in this and subsequent sections is summarized in Table \ref{tab:notation}.

\begin{table}
\begin{tabular}{ l | p{3.6in} l }
  Notation & Definition   & Introduced in\\
 \hline\hline
  $p(x)$     & probability a user leaves the website when inconvenience level is increased by $x$  & Section \ref{sec:users}\\
  $u_0$      & user's initial utility from using the website & Section \ref{sec:ARUM}\\
  $c(x) + Y$ & immediate disutility after an increase in inconvenience by $x$ & Section \ref{sec:ARUM}\\
  $F$        & cumulative distribution function of $Y$ & Section \ref{sec:ARUM}\\
  $A$        & target increase in inconvenience & Section \ref{sec:providers}\\
  $s_A(x)$   & expected fraction of users that use the website after a total increase $A$ is introduced in increments of $x$ & Section \ref{sec:simple}\\
  $l(x)$     & time to adapt completely to a change of magnitude $x$ & Section \ref{sec:time}\\
  $t_A(x)$   & time to introduce a total increase $A$ in increments of $x$ & Section \ref{sec:time}\\
  $\delta$   & provider's discount factor & Section \ref{sec:opt}\\
  $r(x)$     & provider's revenue per user when the total inconvenience is $x$ & Section \ref{sec:opt}\\
  $\Pi(x, z)$& provider's infinite horizon revenue from introducing inconvenience $x$ is each of the next $z$ periods & Section \ref{sec:opt}\\
\hline
\end{tabular}
\vspace{0.25cm}
\caption{Notation used in the paper.}
\label{tab:notation}
\end{table}

\section{Fraction of Users that Stay}
\label{sec:simple}

In this section, we study monotonicity properties of the fraction of users that stay after $A/x$ increases in inconvenience of magnitude $x$ in a simple setting, and in the following sections we discuss under what conditions the result can be generalized.


In this section, we assume that users completely adapt to changes and take the same time to adapt to an increase in inconvenience, independently of the magnitude of the change.  This assumption will be relaxed in the following sections.  We will refer to this adaptation time as ``one period.''

The provider can then introduce the inconvenience $A$ in the following way.  First, increase the inconvenience level by some amount $x$.  Some users will stop using the website because of this inconvenience, but some will stay.  The ones that stay will completely adapt to the change in one period.  Once all remaining users have adapted, the provider can further increase inconvenience by $x$; again, some users will leave, but the ones that stay will adapt one period later.  If the provider repeats this $A/x$ times, the target inconvenience $A$ will be reached, and the expected fraction of users that are still using the website is equal to\footnote{Because $A/x$ is the number of increases, it needs to be an integer.  Thus, the domain of $s_A$ is $\{A/i : i \in \N^+\}$.  However, in our analysis we study monotonicity properties of $s_A$ on $[0,A]$.  Then, monotonicity properties follow for $\{A/i : i \in \N^+\}$.  For instance, if $s_A$ is increasing on $[0,A]$, then it is also increasing in any subset. Note that $s_A(x)$ is well-defined on $[0, A]$ (as long as $p(x)$ is defined).}
\[s_A(x) \equiv p(x)^{A/x}.\]

\begin{figure}
\begin{center}
		\includegraphics[width=0.7\textwidth]{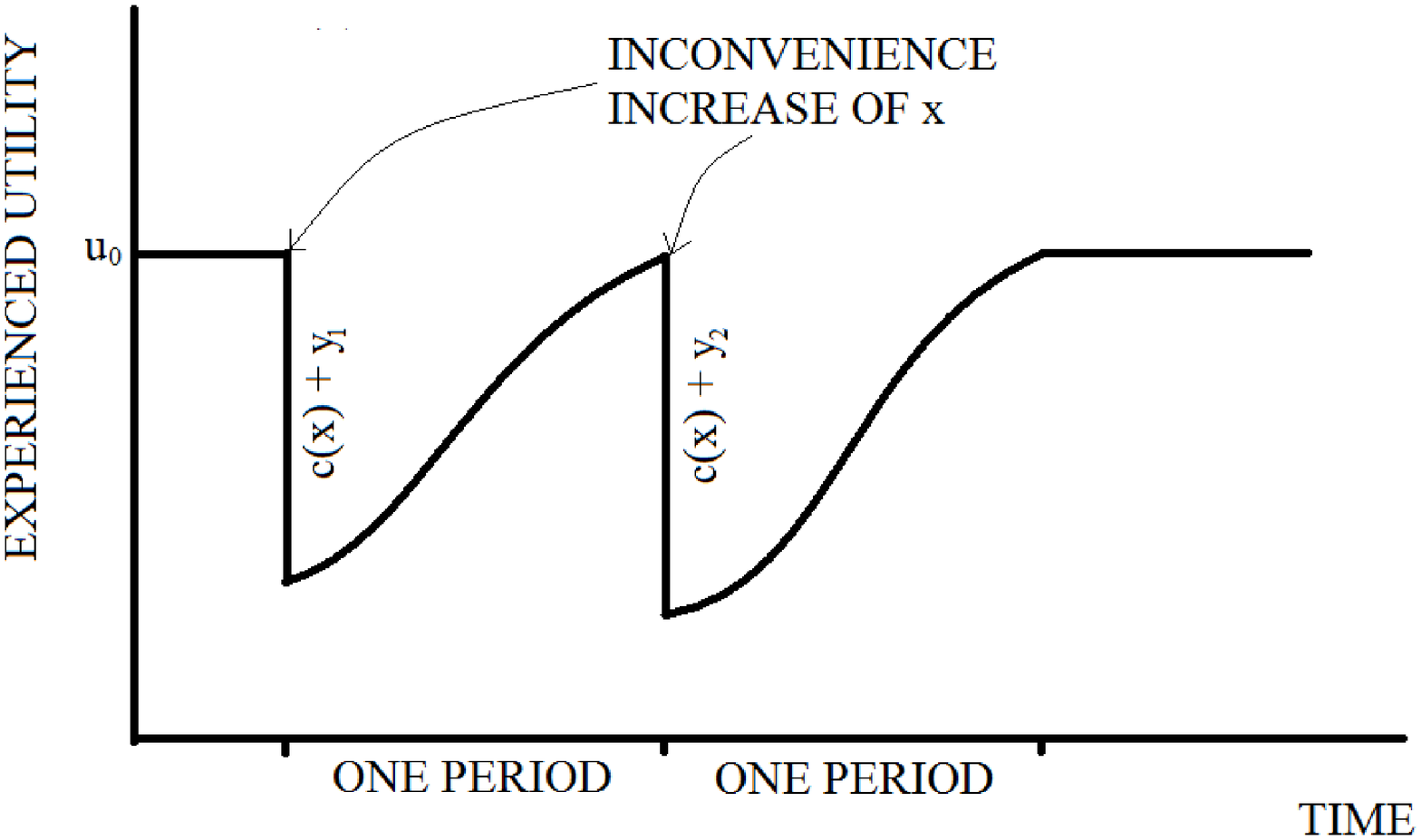}
\end{center}
\vspace{-2ex}
\caption{Schematic utility evolution of a user that stays with the site when the provider implements the change in two steps.  The user stays because his experienced utility never drops below 0.}
\vspace{-1ex}
\label{fig:utility2}
\end{figure}

For example, assume that the provider implements the change in two steps.
Figure \ref{fig:utility2} illustrates the trajectory of a user's utility that continues using the website after both changes have been introduced.  Initially, his utility is equal to $u_0$.  When the first change of magnitude $x$ is introduced, his utility decreases by $c(x) + y_1$, where $y_1$ is the realization of $Y$ after the first increase.  In this case, the disutility is smaller than $u_0$, and thus the user does not leave.  After the change, the user's utility gradually increases (or equivalently, his disutility gradually decreases), and after one period his utility becomes $u_0$.  Then, a second change of magnitude $x$ is introduced, which in this example is the last increase in inconvenience.  Thus, after the user adapts to the second change, his utility remains at $u_0$.  We do not make any assumption on the trajectory of the utility between the time of the change and the time of complete adaptation.

The following lemma gives monotonicity properties of $s_A(x)$.

\begin{lemma}
\label{l:logconcave}
For any $A>0$:
\begin{enumerate}
\item [(i)] If $p(x)$ is log-concave, then
\begin{enumerate}
\item $s_A(x)$ is decreasing in $x$ and
\item for any $x_1,x_2,...,x_z \geq 0$, $p\left(\sum_{j=1}^z x_j\right) \leq \prod_{j=1}^z p(x_j)$.
\end{enumerate}
\item [(ii)] If $p(x)$ is log-convex, then
\begin{enumerate}
\item $s_A(x)$ is increasing in $x$ and
\item for any $x_1,x_2,...,x_z \geq 0$, $p\left(\sum_{j=1}^z x_j\right) \geq \prod_{j=1}^z p(x_j)$.
\end{enumerate}
\end{enumerate}
\end{lemma}

Lemma \ref{l:logconcave} shows that if $p(x)$ is log-concave (resp. log-convex) then the expected fraction of users that use the website after a total increase $A$ is introduced is increasing (resp. decreasing) in the number of changes.  Condition (a) states this directly in terms of $s_A(x)$, that is, assumes that all changes have the same magnitude.  On the other hand, condition (b) is comparing a setting where changes of arbitrary magnitudes are introduced in separate periods with the situation where a change equal to the magnitude of their sums is introduced in one step.
We note that Lemma \ref{l:logconcave} holds regardless of whether $p(x)$ becomes 0 at some finite $x$ or $p(x) > 0$ for all $x$.

We next discuss the assumptions of log-concavity and log-convexity.  Then, in Section \ref{sec:implications} we consider the implications of Lemma \ref{l:logconcave} and relate the result to the ARUM.

\subsection{Log-concave and log-convex functions}
\label{sec:logconcave}

A function is log-concave if its logarithm is concave.  All concave and linear functions are log-concave, but there also exist convex functions that are log-concave.
In this setting, we are interested in whether the function $p(x)$ is log-concave.  This is a decreasing function with $p(0) = 1$ and $p(x) \geq 0$.  Examples of such functions that are log-concave are $e^{-x^k}$ with $k > 1$ and $(1 - x^k) \cdot 1_{\{x \leq 1\}}$ with $k > 1$, where $1_{\{\cdot \}}$ is the indicator function.

A function is log-convex if its logarithm is strictly convex.  Examples of functions that are log-convex and satisfy the requirements of $p(x)$ are $1/(1+x)^k$ with $k > 0$ and $e^{-x^k}$ with $k \in (0, 1)$.

If the function $p(x)$ is differentiable, then log-concavity and log-convexity of $p(x)$ are related to the monotonicity of $p'(x) / p(x)$.  If the ratio $p'(x) / p(x)$ is decreasing (increasing) then $p(x)$ is log-concave (log-convex).


\subsection{Implications of Lemma \ref{l:logconcave}}
\label{sec:implications}

We now consider the implications of Lemma \ref{l:logconcave} for the provider's dilemma.

If $p(x)$ is log-concave, we have the following effects:
\begin{itemize}
\item For a fixed total increase of inconvenience, the faster the final level is reached, the more likely it is that a user leaves the site.  Equivalently, the smaller the number of changes (assuming that all changes have the same magnitude), the more likely a user is to leave the website.
\item A provider that wants to maximize his revenue by increasing some form of inconvenience faces a tradeoff: increasing the inconvenience fast means that he will get higher revenue sooner, but also implies that many existing users will stop using the site.
\end{itemize}

On the other hand, if $p(x)$ is log-convex, the provider does not face a tradeoff.  By increasing the inconvenience in one step, he maximizes the number of users that stay and gets the revenue sooner.\footnote{Proposition \ref{prop:onestep} in Section \ref{sec:opt} shows that if $p(x)$ is log-convex, it is optimal to do the increase in one step for the optimal $A$.  However, this does not hold if the target increase $A$ is very suboptimal.}

\begin{figure}
\begin{center}
		\includegraphics[width=0.5\textwidth]{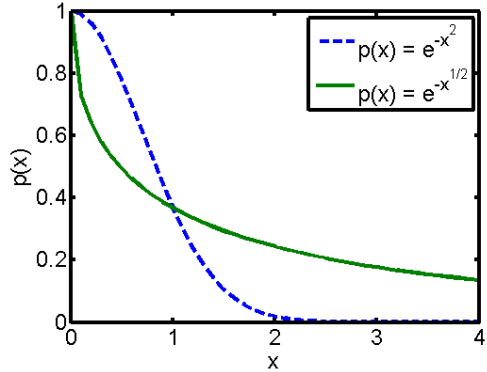}
\end{center}
\vspace{-1ex}
\caption{A log-concave and a log-convex function for the probability of staying after a change of magnitude $x$.  For small changes, more people stay under the log-concave function; for large changes, more people stay under the log-convex function.}
\vspace{-1ex}
\label{fig:comparison}
\end{figure}

We can get some intuition for this result by comparing a log-concave and a log-convex function.  Consider Figure \ref{fig:comparison} which shows the log-concave function $e^{-x^2}$ and the log-convex function $e^{-x^{1/2}}$.  We observe that for small changes, more people stay under the log-concave function.  On the other hand, for large changes, more people stay under the log-convex function.  This suggests that under a log-concave function it is better to make many small changes, whereas under a log-convex function it is better to make one large change.

How does log-convexity of $p(x)$ relate to properties of $c(x)$ and $F(y)$ of the Additive Random Utility Model?
Straightforward calculations show the following lemma.
\begin{lemma}
\label{l:main_ARUM}
If $F$ is log-concave, $F(u_0 - c(0)) = 1$, and $c(x)$ is linear or convex, then $p(x)$ is log-concave.
\end{lemma}

We emphasize that it is the distribution function that is assumed to be log-concave in Lemma \ref{l:main_ARUM}, and not the density function.  In fact, if the density function is log-concave, then the distribution function is also log-concave~\cite{logconcave}.  On the other hand, there exist distributions for which the distribution function is log-concave, while the density function is not (one such example is the log-normal distribution).

Most common distributions are log-concave~\cite{logconcave} (e.g., uniform, normal, exponential, logistic, etc).  However, the assumption $F(u_0 - c(0)) = 1$ can only be satisfied if the support is a subset of $(-\infty, u_0 - c(0)]$.  A log-concave distribution with support $(-\infty, +\infty)$ gives rise to a probability $p(x)$ that is log-concave on $(0, \infty)$ and discontinuous at 0; a property related to the ``penny gap'' phenomenon.  This case is considered in Section \ref{sec:discontinuity}.

Furthermore, the analysis in this section has been based on the following two assumptions:
\begin{enumerate}
\item [(i)] The time to adapt is independent of the magnitude of the change
\item [(ii)] Users adapt to changes completely
\end{enumerate}
In Section \ref{sec:time}, we consider a more general model of complete adaptation, where the time to adapt is an increasing function of the magnitude of the change.  In Section \ref{sec:lasting}, we consider a setting with lasting effects, where users do not completely adapt to changes in finite time.

\section{Discontinuous Behavior}
\label{sec:discontinuity}

In this section, we consider settings of discontinuous behavior: while a user stays with probability 1 when there is no change, a strictly positive change creates a non-negligible probability of leaving no matter how small the change is.
This can be modeled by assuming that the function $p(x)$ is discontinuous at 0; that is, $p(0) = 1$ and $\lim_{x \to 0^+} p(x) < 1$.

\begin{figure}
\begin{center}
		\includegraphics[width=0.5\textwidth]{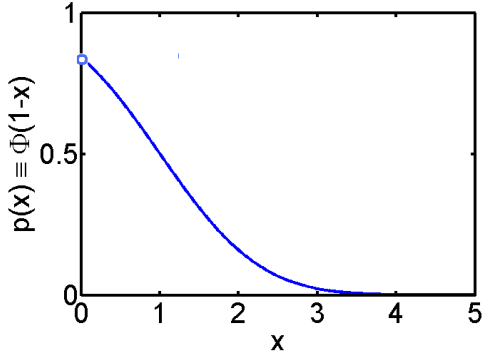}
\end{center}
\vspace{-1ex}
\caption{Plot of $p(x)$ for $x > 0$ when $Y \sim N(0,1)$ and $c(x) = x$.  In this case, $\lim_{x \to 0^+} p(x) < 1$.}
\vspace{-3ex}
\label{fig:linear_cost}
\end{figure}

In fact, in the ARUM model, if the noise $Y$ is drawn from a distribution whose support includes the interval $(a, \infty)$ for some constant $a$, then $\lim_{x \to 0^+} p(x) < 1$.  By definition, $p(0) = 1$, and thus we have a discontinuity at zero.
For illustration, Figure \ref{fig:linear_cost} shows $p(x)$ for $u_0 = 1$ and $c(x) = x$ when $Y$ is drawn from the standard normal distribution.
We observe that there is a discontinuity at 0, since $p(0) = 1$.  We note that the gap $p(0) - \lim_{x \to 0^+} p(x)$ decreases as $u_0$ increases, but never becomes equal to 0.\footnote{We can only have continuity at 0 if $F(u_0 - c(0)) = 1$.  This can only be the case if the maximum point of the distribution's support is finite and equal to $u_0 - c(0)$ (see Lemma \ref{l:main_ARUM}).}

We note that $p(x)$ can be discontinuous at 0 and log-convex throughout $[0, \infty)$.  Then, according to Lemma \ref{l:logconcave}, $s_A(x)$ is increasing in $x$.
On the other hand, $p(x)$ cannot be discontinuous at 0 and log-concave throughout $[0, \infty)$.  The following proposition considers functions that are discontinuous at 0 and log-concave on $(0, \infty)$.  It is shown that the corresponding $s_A(x)$ is unimodal: increasing for small values of $x$ and decreasing for large values of $x$.

\begin{proposition}
\label{prop:discont}
If $\lim_{x \to 0^+} p(x) < p(0) \equiv 1$, and $p(x)$ is log-concave for $x > 0$, then there exists $\bar{x} > 0$ such that $s_A(x)$ is increasing for $x \in (0, \bar{x})$ and decreasing for $x \in (\bar{x}, \infty)$.
\end{proposition}

Thus, when $p(x)$ is log-concave and discontinuous at zero, then $s_A(x)$ is not decreasing throughout its domain.  In the extreme case, $\bar{x} > A$ and thus $s_A(x)$ is increasing in $[0, A]$.  For instance, if $A = 0.5$ and $p(x) = 0.5 e^{-x^2}$, then $s_A(x)$ is increasing in $[0, A]$.
\comment{
In this extreme case, even though $p(x)$ is log-concave on $(0, \infty)$, it is optimal for the provider to introduce the optimal target increase in one step.  However, in most cases, it will be optimal to introduce the inconvenience gradually when $p$ is log-concave on $(0, \infty)$ --- even in the presence of a zero-price effect.
}

\begin{figure}
\begin{center}
		\includegraphics[width=0.7\textwidth]{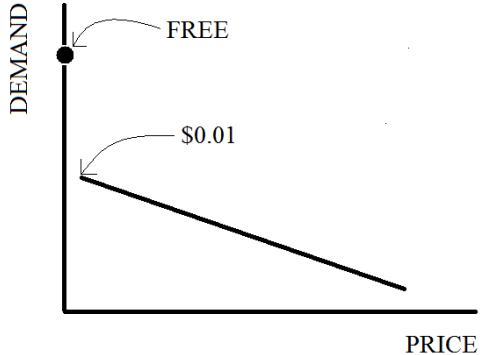}
\end{center}
\vspace{-4ex}
\caption{The penny gap phenomenon.  Observe the discontinuity in the demand when the price is 0.}
\vspace{-1ex}
\label{fig:discont}
\end{figure}

The discontinuous behavior considered in this section is related to discontinuous behaviors in other settings, most notably, those involving a zero price.
There is considerable empirical evidence that decisions about free content and products differ from those involving a  price, however small. Specifically,  the benefits associated with free content are perceived to be higher than those with even minimal cost, which implies that when confronted with free choices benefit-cost analyses are not part of the decision process. Thus people are much more likely to take a product that is given to them for free than to choose something better at a very low price.
This phenomenon, called the {\em zero-price effect}~\cite{ariely:zero} or the {\em penny gap}~\cite{free, penny_gap}, has also been observed in the context of information technology, where affordable content turns out to be much harder to provide than free one~\cite{NYT:content}.
Formally, the zero-price effect implies that the demand is discontinuous at zero, as illustrated in Figure \ref{fig:discont}~\cite{penny_gap}.
We note, however, that the setting considered is this section does not correspond to the zero-price effect, since even if the inconvenience that is introduced consists of a subscription cost for content that was initially free, our model assumes discontinuous behavior every time that price increases (not only at the first period when the price increases from zero to some positive amount).

\section{Time to Adapt}
\label{sec:time}

The previous sections assumed that the time to adapt to any change is constant and independent of the magnitude of the change.  Here we relax this assumption.

Let $l(x)$ be the time it takes a user to completely adapt to a change of magnitude $x$.\footnote{We remind the reader that according to our ARUM, the user incurs a cost of $c(x)+Y$ after an increase in inconvenience by $x$, where $Y$ is a random variable.  Thus, by denoting the time to adapt by $l(x)$, we are implicitly assuming that it does not depend on $Y$, which may seem unrealistic.  To address this, we can consider a more general model where the time to adapt is $\tilde{l}(x, y)$, where $x$ is the magnitude of the change and $y$ is the realization of the random variable $Y$.  Assuming that $\tilde{l}$ is increasing in $y$, we define $l_{\epsilon}(x) = \tilde{l}(x, F^{-1}(1 - \epsilon))$ for $\epsilon \geq 0$.  If $F$ has a finite support, then in time $l_0(x)$ every user will adapt to a change of magnitude $x$, so we can set $l(x) = l_0(x)$.  Otherwise, if $F$ has an infinite support (or if $F$ has a finite support but $l_0(x)$ is very large), we can choose some $\epsilon > 0$ and set $l(x) = l_{\epsilon}(x)$.  In this case, $(1-\epsilon) p(x)$ is a lower bound for the probability that a user stays; by replacing $p(x)$ by $(1-\epsilon) p(x)$ we can perform a worse-case analysis.  The provider could optimally choose $\epsilon$; however, this is beyond the scope of this paper. }
We expect this to be a nondecreasing function: the greater the increase (i.e., $x$) the more time it takes to adapt.  In the terminology of Section \ref{sec:simple}, $l(x)$ is the duration of a period.
Now the provider has to wait for a time $l(x)$ between any two consecutive increases in order to give users that continue using the website enough time to adapt to the change.  Thus, the total time it takes for the website to make a total change $A$ through $A/x$ changes of size $x$ is
\[t_A(x) \equiv \left(\frac{A}{x}-1\right) l(x).\]

In the previous section we considered the special case where $l(x)$ is constant (and positive) for $x>0$.
We saw that the provider faces a tradeoff if $p(x)$ is log-concave: a smaller increase $x$ per period implies that more users will adapt and continue using the site, but it will take more time until the target level of inconvenience is reached.  We now generalize this result for a wider class of functions $l$.

\begin{proposition}
\label{prop:tradeoff}
If $p(x)$ is log-concave, and
\begin{equation}
\label{eq:elasticity}
x \cdot l'(x) < l(x) \text{ for } x>0,
\end{equation}
then
\begin{enumerate}
\item [(i)] Both $s_A(x)$ and $t_A(x)$ are decreasing in $x$.
\item [(ii)] The expected number of users after $A/x$ increases of size $x$ is a decreasing function of the average rate of increase $\bar{r}$, where $\bar{r} \equiv A/t_A(x)$.
\end{enumerate}

\end{proposition}

We note that $x \cdot l'(x) / l(x)$ is the elasticity of $l$.  Thus, \eqref{eq:elasticity} requires that the time to adapt is inelastic in the change.  The case that $l$ is constant (considered in Section \ref{sec:simple}) is a special case where the elasticity is equal to 0.

Proposition \ref{prop:tradeoff} says that if the probability of staying after a change is log-concave in the magnitude of the change and the time to adapt to a change is inelastic in the magnitude of the change, then
\begin{itemize}
\item The provider faces a tradeoff between achieving a higher revenue per user sooner and maximizing the number of users in the long term.
\item For a fixed target level of inconvenience, the long-term number of users is a decreasing function of the average rate of increase.
\end{itemize}

If $p(x)$ is log-concave and $l(x)$ is elastic, then the provider again faces a tradeoff, but not with the monotonicity properties discussed above, since the function $t_A(x)$ is not decreasing for all $x > 0$.  
On the other hand, if $p(x)$ is log-convex, then by selecting $x = A$ (i.e., performing the change in one step) both the expected fraction of long-term users ($s_A(x)$) is maximized and the time until the total time is minimized (because $t_A(A) = 0$).

\section{Incomplete Adaptation}
\label{sec:lasting}

Up to now we have assumed that users that stay after a change have adapted completely by the time of the next change.  As a result, after every change of magnitude $x$, a user stays with probability $p(x)$.

In this section, we consider the case of lasting effects.  We assume that users do not completely adapt after a change, and as a result {\em the probability of staying after a change may be a decreasing function of the total inconvenience introduced thus far}.  To model this, let $p_i(x)$ be the probability that a user stays when the $i$-th change of magnitude $x$ is introduced (assuming that he stayed after all previous changes).  Under complete adaptation $p_i(x)$ is simply $p(x)$, that is, independent of $i$; under incomplete adaptation $p_i(x)$ will be decreasing in $i$.

The probability that a user stays after the change of $A$ is completed in increments of $x$ is
\[s_{A, \epsilon} (x) \equiv \prod_{i=1}^{A/x} p_i(x).\]
Note that $s_{A, \epsilon = 0}(x) = s_A(x)$.
Thus, we know from Lemma \ref{l:logconcave} that if $p(x)$ is log-concave, then $s_{A, \epsilon = 0}(x)$ is decreasing in $x$.  This is the case of complete adaptation.

We can model lasting effects either directly through the probability $p(x)$ or through the ARUM.  These approaches are considered in parts (i) and (ii) of the following proposition.

\begin{proposition}
\label{prop:lasting}
Suppose that $p(A/k) > 0$ for some $k \in \N$.  Let $d(\cdot)$ be a strictly increasing function such that $d(0) = 0$, and let $\epsilon > 0$.
If either of the following hold:
\begin{enumerate}
\item [(i)] $p_i(x) \equiv p(x) - \epsilon \cdot d((i-1) x)$
\item [(ii)] The ARUM applies and once a user adapts to the $i$-th change of magnitude $x$ his utility is equal to $u_0 - \epsilon \cdot d(i \cdot x)$, and $F(u_0 - \tilde{\epsilon}) < 1$ for any $\tilde{\epsilon} > 0$
\end{enumerate}
then there exist $x_1$, $x_2$ such that $x_1 < x_2$ and $s_{A, \epsilon}(x_1) < s_{A, \epsilon}(x_2)$.
\end{proposition}

The function $d$ represents how much the probability of staying is decreased because of the total inconvenience that the user is currently experiencing.
In the ARUM case (case (ii)), this effect is caused from a permanent decrease in the utility function.  In particular, in the ARUM case, the function $d$ represents how much the user's utility is permanently decreased because of the increases in inconvenience.  We are assuming that the decrease in the user's utility is a function of the inconvenience that the user is currently experiencing, that is, the sum of all inconveniences introduced up to the present time.
The constant $\epsilon$ represents the magnitude of the lasting effect.

We have previously seen that under complete adaptation and log-concave probabilities of staying, the long-term number of users satisfies a monotonicity property: it is an increasing function of the number of periods it takes to perform the change.  (This is the context of Lemma \ref{l:logconcave}.)
Proposition \ref{prop:lasting} shows that even a small lasting effect destroys this monotonicity.  The reason is that no matter how small the lasting effect is, it accumulates over a large number of periods.  As a result, the expected number of users decreases much more if the change is performed through many small increases than if the change is performed through a few increases of a larger magnitude.

\section{Revenue Maximization}
\label{sec:opt}

In all previous sections, we were assuming that the target increase in inconvenience $A$ was fixed and studied monotonicity properties of $s_A(x)$, which is the expected fraction of long-term users when $A$ is introduced in increments of $x$.  In this section we consider the problem of maximizing the expected discounted payoff of the provider over both $x$ and $A$ (or equivalently $x$ and $z \equiv A/x$).

We denote the revenue per user from an inconvenience of $x$ by $r(x)$.  We assume that $r(x)$ is an increasing function: the higher the inconvenience to the user, the higher the revenue to the provider.  If this were not true, the provider would decrease the inconvenience to make both himself and the users better off.

We assume that each user visits the website once every period.  The provider discounts future payments according to a discount factor $\delta$.
We further assume that an increase in inconvenience is made only after users that continued using the site after the previous increase have completely adapted.

If we assume that the magnitude of all increases is the same and that users that stay adapt to changes in one period, then the provider needs to solve the following problem:

\[ \max_{x \geq 0, z \in \N^+}   \Pi(x, z) \equiv \sum_{i=1}^{z-1} \delta^{i-1} p(x)^i r(x \cdot i) + \frac{\delta^{z-1}}{1 - \delta} p(x)^z r(x \cdot z) \]

In particular, $\Pi(x, z)$ denotes the provider's infinite horizon revenue from introducing inconvenience $x$ is each of the next $z$ periods.
After the $i$-th increase, a user is still using the system with probability $p(x)^i$ and the provider gets a revenue of $r(x \cdot i)$ per user in that period which he discounts by $\delta^{i-1}$.

The problem of maximizing $\Pi(x, z)$ is two-dimensional, since we wish to choose both $x$ and $z$ optimally.
In the following section we show that if $p$ is log-convex, then it is optimal to implement the change in one step (validating the claim made in Section \ref{sec:simple}).  Then, we show that for any $p$, the problem can be reduced to a one-dimensional problem if $r$ is log-concave.

\subsection{Log-convex Probability of Staying}
\label{sec:log-convex}

In this section we consider the case of a log-convex $p(x)$.   We have seen in Lemma \ref{l:logconcave} that if $p(x)$ is log-convex then $s_A(x)$ is increasing in $x$, which implies that the long-term number of users is maximized when the increase is performed faster.  The following proposition shows that $\Pi(x, z)$ is maximized when the increase is performed in one step (i.e., $z=1$).  As a result, the problem of maximizing $\Pi(x, z)$ reduces to maximizing $p(x) \cdot r(x)$.

\begin{proposition}
\label{prop:onestep}
Suppose $p(x)$ is log-convex.  Let $x^* \in \arg \max_{x \geq 0} \{p(x) \cdot r(x)\}$.  Then $(x^*, 1)$ is a maximizer of $\Pi(x, z)$.
\end{proposition}

The previous proposition shows that $(x^*, 1)$ is always a maximizer of $\Pi(x, z)$.  We note that if $p(x)^i \cdot r(i \cdot x) =  p(x)^z \cdot r(z \cdot x)$ for $i=1, 2, ..., z-1$ and $p(x)^z = p(x \cdot z)$, then there may exist other maximizers as well.   However, this is not the case if $r$ is strictly log-convex.

We note that the proof of Proposition \ref{prop:onestep} can be extended to show that if $x_i$ is the magnitude of the increase in period $i$, then to find the optimal solution it suffices to maximize $p(x) \cdot r(x)$ and implement the increase in one period.

Proposition \ref{prop:onestep} shows that the optimal solution is very simple if $p$ is log-convex: the provider only needs to maximize $p(x) \cdot r(x)$ and perform the optimal increase right away.  However, we expect that the function $p(x)$ will usually not be log-convex.  In particular, log-convexity is associated with a non-negligible probability of staying when the magnitude of the inconvenience is very large, which is often not the case.  On the other hand, the fact that there are no standard distribution functions that are log-convex implies that log-convexity of $p(x)$ is highly unlikely if $p(x)$ comes from the ARUM.

Thus, even though in the case of log-convexity the provider's optimal decision is straightforward, we expect that usually he will face a tradeoff between maximizing the number of long-term users and minimizing the time.  This problem is studied in the following section.

\subsection{Log-concave Revenue per User}
\label{sec:revenue}

In this section, we assume that $r$ is log-concave.  As mentioned in Section \ref{sec:logconcave}, the class of log-concave functions includes all concave and linear functions.  These are reasonable assumptions for a revenue function, because such functions exhibit constant or decreasing marginal returns.  Because of the generality of log-concavity, our results apply to a variety of situations.
When the inconvenience is generated by a subscription cost, then the revenue per user is equal to the subscription fee itself, and thus $r(x) = x$.  On the other hand, when the inconvenience is because of advertising, then $r(x)$ can model various pricing schemes for online advertising (e.g., pricing per impression, pricing per click, and pricing per acquisition).  Moreover, the price-per-impression and the price-per-click could either be exogenously defined or depend on the total number of advertisements on the site.

We next show how for any fixed magnitude of inconvenience $x$ we can find the optimal number of times that an inconvenience of magnitude $x$ should be introduced.
We denote this number by $z^*(x)$.

\begin{lemma}
\label{l:z*}
If $r$ is log-concave, then for a fixed $x$, $\Pi(x, z)$ is maximized at
\[z^*(x) = \min\left\{z \in \N: \frac{r(x \cdot z)}{r(x \cdot (z+1))} \geq p(x)\right\}.\]
\end{lemma}

We can get some intuition for this result by considering that after the $z$-th change is introduced, the provider gets $r(x \cdot z)$ from each user per period.  Increasing the inconvenience by $x$ one more time will result in a revenue of $r(x \cdot (z+1))$ from each remaining user and each user will stay with probability $p(x)$.  Thus the change is worthwhile if and only if $r(x \cdot z) \leq p(x) \cdot r(x \cdot (z+1))$.  Because $r$ is log-concave, the ratio $r(x \cdot z) / r(x \cdot (z+1))$ is increasing in $z$, which implies that it is never profitable to increase $z$ above $z^*(x)$.


\begin{figure}
\begin{center}
		\includegraphics[width=0.5\textwidth]{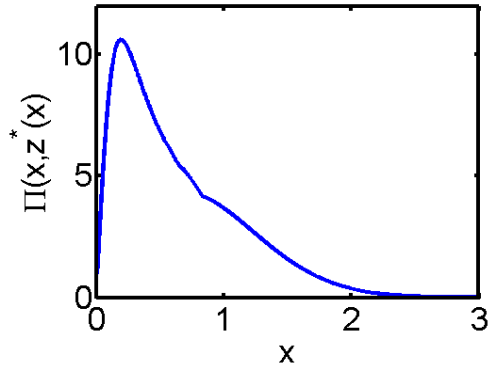}
\end{center}
\vspace{-2ex}
\caption{$\Pi(x, z^*(x))$ for Example \ref{ex:opt}.  It is maximized at $0.195$, which implies that the revenue is maximized if the inconvenience increases $z^*(0.195) = 26$ times by an amount of $0.195$}
\vspace{-1ex}
\label{fig:example}
\end{figure}

Lemma \ref{l:z*} reduces the problem of maximizing $\Pi(x, z)$ to the one-dimensional problem of maximizing $\Pi(x, z^*(x))$.
Note that Lemma \ref{l:z*} does not make any assumptions on $p(x)$.  It applies for any $p(x)$, whether it is log-concave, log-convex, or neither; discontinuous at 0; etc.  However, since we already know (from Proposition \ref{prop:onestep}) how to maximize $\Pi(x, z)$ when $p(x)$ is log-convex, Lemma \ref{l:z*} will be useful when $p(x)$ is not log-convex.

The following example applies Lemma \ref{l:z*} to maximize the provider's revenue for an instance of the problem.

\begin{example}
\label{ex:opt}
Suppose $\delta = 0.9$, $p(x) = e^{-x^2}$ and $r(x) = x$.  We plot $\Pi(x, z^*(x))$ in Figure \ref{fig:example}.  At the optimal solution $(x, z) = (0.195, 26)$, that is, the revenue is maximized if the inconvenience increases $26$ times by an amount of $0.195$.  Thus, if $p(x)$ represents the probability of staying when the subscription fee is increased by $x$ dollars, then it is optimal to increase the subscription fee by about \$0.20 for a total of 26 times until reaching a final subscription fee of approximately \$5.20.
\end{example}

\subsubsection*{Lasting effects}

Lemma \ref{l:z*} can be generalized to include lasting effects.

\begin{lemma}
Suppose that a user stays after the $i$-th increase of magnitude $x$ with probability $p(x) - \epsilon \cdot (i-1) \cdot x$.
If $r$ is log-concave, then for a fixed $x$, $\Pi(x, z)$ is maximized at
\[z^*(x; \epsilon) = \min\left\{z \in \N: \frac{r(x \cdot z)}{r(x \cdot (z+1))} + \epsilon \cdot d(z\cdot x) \geq p(x)\right\}.\]
\end{lemma}
Note that for $\epsilon = 0$ this is identical to Lemma \ref{l:z*}.
We observe that $z^*(x; \epsilon)$ is non-increasing in $\epsilon$ for a fixed $x$.

\section{Conclusion}
\label{sec:conclusion}

This paper studies revenue maximization from the point of view of an established content provider with an existing user base through the lens of adaptation theory.  The provider can increase revenues by imposing some inconvenience to users while risking to lose some of the users.  Our approach is very general in that it can be applied for any  revenue generating process that imposes inconvenience to the users (e.g., advertisements, subscription fees).  

Our analysis is based on the function $p(x)$ that represents the probability that a user stays after an increase in inconvenience of magnitude $x$.
We provide a utility model from which $p(x)$ may arise; however, knowledge of the utility model is not essential for applying the results.  In particular, the provider can directly use $p(x)$ to find the optimal strategy that maximizes his revenue.

The provider can use A/B testing to estimate $p(x)$.  For a given value of $x_i$, the provider can impose this inconvenience to some users.  The percentage of these users that continue using the website is an estimate for $p(x_i)$.  The provider should only use a small percentage of users to estimate $p(x)$.  Once the provider has a good estimate for $p(x)$ through which he can compute the optimal way to introduce the inconvenience, then the optimal inconvenience is applied to all users.  We note that $p(x)$ can also be estimated from information from past experience and surveys.



\newpage
\bibliographystyle{abbrv}
\bibliography{ref}

\include{appendix}

\end{document}

%% file: appendix.tex
\section*{Appendix}

\noindent  \textbf{{\em Proof of Lemma~\ref{l:logconcave}:}}
Let $g(x) = \log(p(x))$.
Since $A$ is a constant, $s_A(x)$ has the same monotonicity properties as
\[h(x) \equiv \log((p(x))^{1/x}) = \frac{g(x)}{x}.\]
We first show (i).  If $p(x)$ is log-concave, then $g(x)$ is concave.  Let $x_1 < x_2$.  Concavity of $g$ implies that for any $x_2 > 0$ and $\theta \in (0, 1)$
\[g(\theta x_2) \geq (1 - \theta) g(0) + \theta g(x_2).\]
Note that $g(0) = 0$, because $p(0) = 1$.  Setting $\theta = x_1/x_2$ implies that
\[\frac{g(x_1)}{x_1} \geq \frac{g(x_2)}{x_2},\]
which shows that $s_A(x)$ is decreasing in $x$.

If $\log(p(x))$ is concave, Jensen's inequality implies that
\[\log \left(p\left(\frac{1}{z} \sum_{j=1}^z x_j\right)\right) \leq \frac{1}{z} \sum_{j=1}^z \log (p(x_j)).\]
Moreover, concavity of $\log(p(x))$ and the fact that $\log(p(0)) = 0$ imply that
\[z \log \left(p\left(\frac{1}{z} \sum_{j=1}^z x_j\right)\right) \leq \log \left(p\left(\sum_{j=1}^z x_j\right)\right).\]
Combining the last two inequalities we get that
\[p\left(\sum_{j=1}^z x_j\right) \leq \prod_{j=1}^z p(x_j).\]

A similar argument shows (ii).
\qed

\vspace{0.5cm}

\begin{figure}
\begin{center}
		\includegraphics[width=0.7\textwidth]{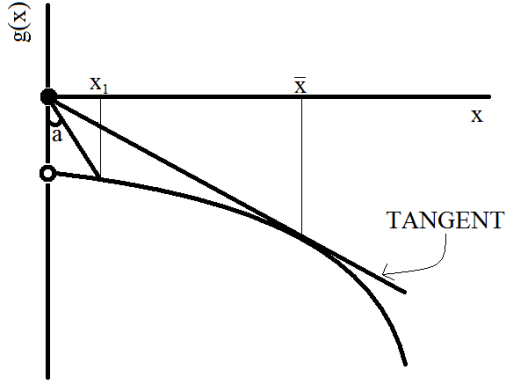}
\end{center}
\vspace{-2ex}
\caption{The function $g(x) \equiv \log(p(x))$ and its tangent at $\bar{x}$.  The cotangent of the angle $a$ is equal to $g(x_1) / x_1$.}
\vspace{-1ex}
\label{fig:tangent}
\end{figure}

\noindent \textbf{{\em Proof of Proposition~\ref{prop:discont}:}}
Let $g(x) = \log(p(x))$.  By the assumptions of this proposition, $g$ is concave, $g(0) = 0$ and $\lim_{x \to 0^+} g(x) < 0$.
Let $\bar{x}$ be such that $x \cdot g(\bar{x})/\bar{x}$ is tangent to $g(x)$.  This is shown schematically in Figure \ref{fig:tangent}.
It suffices to show that $g(x)/x$ is increasing for $x \in (0, \bar{x})$ and decreasing for $x \in (\bar{x}, \infty)$, because $\log(s_A(x)) = A g(x) / x$.

Consider some $x_1 < \bar{x}$.  We observe in Figure \ref{fig:tangent} that $g(x_1)/x_1$ is equal to the cotangent of angle $a$.  The angle increases as $x_1$ increases (as long as $x_1 < \bar{x}$).  Since the cotangent decreases in $(0^{\circ}, 90^{\circ})$, we conclude that if $x_1 < x_2 < \bar{x}$ then $g(x_1)/x_1 \geq g(x_2)/x_2$.

On the other hand, to see why $s_A(x)$ is decreasing in $x$ for $x > \bar{x}$, define
\[\tilde{g}(x) = x \cdot \frac{g(\bar{x})}{\bar{x}} \cdot 1_{\{x < \bar{x}\}} + g(x) \cdot 1_{\{x \geq \bar{x}\}}.\]
The function $\tilde{g}$ is concave in $[0, \infty)$ and $\tilde{g}(0) = 0$, so the same argument as Lemma \ref{l:logconcave} shows that $\tilde{g}(x)/x$ is decreasing in $x$.  Because of the way we defined $\tilde{g}$, $\tilde{g}(x) = g(x)$ for $x \geq \bar{x}$.  This implies that $g(x)/x$ is decreasing for $x \in (\bar{x}, \infty)$.
\qed

\vspace{0.5cm}

\noindent \textbf{{\em Proof of Proposition~\ref{prop:tradeoff}:}}
We first show that if $x \cdot l'(x) < l(x)$, then $t_A(x)$ is decreasing.
\begin{align*}
t_A'(x) &= l'(x) \left(\frac{A}{x}-1\right) - l(x) \frac{A}{x^2} \\
      &= \frac{l(x)}{x} \left(\frac{x \cdot l'(x)}{l(x)} \left(\frac{A}{x}-1\right) - \frac{A}{x}\right) \\
      &< \frac{l(x)}{x} \left( \left(\frac{A}{x}-1\right) - \frac{A}{x}\right) \\
      &< 0
\end{align*}
Now part (i) follows from Lemma \ref{l:logconcave}.

We next prove part (ii).  Let
\[f_A(x) \equiv \frac{A}{t_A(x)}\]
We have shown that $t_A$ is decreasing, which implies that $f_A$ is increasing.  Thus, the inverse $f_A^{-1}$ is an increasing function.
Since the average rate of increase $\bar{r}$ is equal to $f_A(x)$, we conclude that the increase $x$ satisfies
$x = f_A^{-1}(\bar{r})$.
The probability that a user continues using the website after $A/x$ increases of $x$ is $s_A(x) \equiv p(x)^{A/x}$, and as a function of $\bar{r}$ it can be expressed as $s_A(f_A^{-1}(\bar{r}))$, which by Lemma \ref{l:logconcave} is a decreasing function of $\bar{r}$.
\qed

\vspace{0.5cm}

\noindent \textbf{{\em Proof of Proposition~\ref{prop:lasting}:}}
We first show (i).
\begin{align*}
s_{A, \epsilon} (x)
     & = \prod_{i = 1}^{A/x} (p(x) - \epsilon \cdot d((i-1) x)\\
     & \leq \prod_{i=A/(2x)+1}^{A/x} (p(x) - \epsilon \cdot d((i-1) x)\\
     & \leq (p(x) - \epsilon \cdot d(A/2))^{A/(2x)}\\
     & \leq (1 - \epsilon \cdot d(A/2))^{A/(2x)}\\
\end{align*}
Since $1 - \epsilon \cdot A/2 < 1$, for any $\epsilon, \delta > 0$ there exists a $\bar{x} > 0$ such that $s_{A, \epsilon} (x) < \delta$ for $x \in (0, \bar{x})$.  In particular, this is achieved for $\bar{x}(\delta, \epsilon) = A \log(1 - \epsilon d(A/2)) / \log(\delta)$.  The result follows by choosing some $x_2$ with $p(x_2) > 0$, setting $\delta = p(x_2)$ and $x_1 < \min(x_2, \bar{x}(\delta, \epsilon))$.

We now show (ii).  We observe that if (ii) holds, then the probability that a user stays after the $i$-th increase (given that he stayed after all previous increases) is
\begin{align*}
p_i(x) &= \P[Y < u_0 - c(x) - \epsilon d((i-1) \cdot x)] \\
                      &= F(u_0 - c(x) - \epsilon d((i-1) \cdot x)).
\end{align*}
Thus,
\begin{align*}
s_{A, \epsilon} (x)
     & = \prod_{i=1}^{A/x} F(u_0 - c(x) - \epsilon \cdot d((i-1) x))\\
     & \leq \prod_{i=A/(2x)}^{A/x-1} F(u_0 - c(x) - \epsilon \cdot d((i-1) x))\\
     & \leq (F(u_0 - c(x) - \epsilon \cdot d(A/2)))^{A/(2x)}\\
     & \to 0 \text{ as } x \to 0
\end{align*}
Since $u_0 - c(x) < 1$, for any $\epsilon, \delta > 0$ there exists a $\bar{x} > 0$ such that $s_{A, \epsilon} (x) < \delta$ for $x \in (0, \bar{x})$.  The result follows by choosing some $x_2$ with $p(x_2) > 0$, setting $\delta = p(x_2)$ and $x_1 < \bar{x}(\delta, \epsilon)$.
\qed

\vspace{0.5cm}

\noindent \textbf{{\em Proof of Proposition~\ref{prop:onestep}:}}
Suppose there exists an optimal solution $(x,z)$ with $z > 1$ that is strictly better than $(x^*, 1)$.  We will contradict this by showing that either $\Pi(x \cdot z, 1) \geq \Pi(x, z)$ or $\Pi(x \cdot z', 1) > \Pi(x, z)$ for some $z' < z$.  This will imply that there is a solution $(x', 1)$ that is at least as good as $(x, z)$, and thus $\Pi(x^*, 1) \geq \Pi(x', 1) \geq \Pi(x, z)$.

Since $p(x)$ is log-convex and $z > 1$, we know (from Lemma \ref{l:logconcave}) that $p(x)^{1/x} \leq p(x \cdot z)^{1/(x \cdot z)}$ which implies that
\begin{equation}
\label{eq:p-log-conv}
p(x)^z \leq p(x \cdot z).
\end{equation}

First suppose that $p(x)^i \cdot r(i \cdot x) \leq  p(x)^z \cdot r(z \cdot x)$ for $i=1, 2, ..., z-1$.  Then, by \eqref{eq:p-log-conv} we have that $p(x)^i \cdot r(i \cdot x) \leq  p(x \cdot z) \cdot r(z \cdot x)$ for $i=1, 2, ..., z$.  Thus, in this case,
\begin{align*}
\Pi(x, z) &= \sum_{i=1}^{z-1} \delta^{(i-1)} p(x)^i r(i x) + \frac{\delta^{(z-1)}}{1 - \delta} p(x)^z r(x \cdot z)\\
          &\leq \sum_{i=1}^{z-1} \delta^{(i-1)} p(x \cdot z) r(x \cdot z) + \frac{\delta^{(z-1)}}{1 - \delta} p(x \cdot z) r(x \cdot z)\\
          &= \frac{1}{1 - \delta} p(x \cdot z) r(x \cdot z)\\
          &= \Pi(x \cdot z, 1)
\end{align*}

Now suppose that $p(x)^i \cdot r(i \cdot x) > p(x)^z \cdot r(z \cdot x)$ for some $i < z$.  Let
\[z' = \min \{i : i \in \arg \max \{p(x)^i \cdot r(i \cdot x)\}\}\]
Then,
\begin{align*}
\Pi(x, z) &= \sum_{i=1}^{z-1} \delta^{(i-1)} p(x)^i r(i x) + \frac{\delta^{(z-1)}}{1 - \delta} p(x)^z r(x \cdot z)\\
          &< \sum_{i=1}^{z'-1} \delta^{(i-1)} p(x \cdot z) r(x \cdot z) + \frac{\delta^{(z'-1)}}{1 - \delta} p(x)^{z'} r(x \cdot z')\\
          &\leq \frac{1}{1 - \delta} p(x \cdot z') r(x \cdot z')\\
          &= \Pi(x \cdot z', 1)
\end{align*}
\qed

\vspace{0.5cm}

\noindent \textbf{{\em Proof of Lemma~\ref{l:z*}:}}
\begin{align*}
&\Pi(x, z+1) - \Pi(x, z)  = \\
&\sum_{i=1}^{z} \delta^{(i-1)} p(x)^i r(i x) + \frac{\delta^z}{1 - \delta} p(x)^{z+1} r((z+1) \cdot x) - \sum_{i=1}^{z-1} \delta^{(i-1)} p(x)^i r(i x) - \frac{\delta^{z-1}}{1 - \delta} p(x)^z r(z \cdot x) = \\
&\delta^{(z-1)} p(x)^z r(z x) - \frac{\delta^{z-1}}{1 - \delta} p(x)^z (r(z \cdot x) - \delta p(x) r((z+1) \cdot x)) = \\
&\delta^{(z-1)} p(x)^z (r(z x) - \frac{1}{1 - \delta} r(z \cdot x) + \frac{\delta}{1 - \delta} p(x) r((z+1) \cdot x)) = \\
&\frac{\delta^z}{1 - \delta} p(x)^z (p(x) r((z+1) \cdot x) - r(z x))
\end{align*}
Thus, $\Pi(z+1, x) > \Pi(z, x)$ if $p(x) > r(z x) / r((z+1) \cdot x)$, and $\Pi(z+1, x) < \Pi(z, x)$ if $p(x) < r(z x)/ r((z+1) \cdot x)$.  Moreover, since $r$ is log-concave, $r(z x)/ r((z+1) \cdot x)$ is increasing in $z$ (for a fixed $x$). Thus, $\Pi(z, x)$ is unimodal in $z$ for a fixed $x$: it is increasing for $z < z^*(x)$ and decreasing for $z > z^*(x)$.  It is thus maximized at $z^*(x)$.
\qed

%% file: rate_adaptation.bbl
\begin{thebibliography}{10}

\bibitem{LondonTimes}
{London Times} readership drops 90\% and the {New York Times} could be next.
\newblock {\em InvestorPlace}, July 22, 2010.

\bibitem{free}
C.~Anderson.
\newblock {\em Free: The Future of a Radical Price}.
\newblock Hyperion, 2009.

\bibitem{logconcave}
M.~Bagnoli and T.~Bergstrom.
\newblock Log-concave probability and its applications.
\newblock {\em Economic Theory}, 26(2):445--469, 2005.

\bibitem{baye:ad}
M.~R. Baye and J.~Morgan.
\newblock A simple model of advertising and subscription fees.
\newblock {\em Economics Letters}, 69:345--351, 2000.

\bibitem{brand:free_info}
S.~Brand.
\newblock {\em The Media Lab: Inventing the Future at M.I.T.}
\newblock Penguin, 1988.

\bibitem{microeconometrics}
A.~C. Cameron and P.~K. Trivedi.
\newblock {\em Microeconometrics: Methods and Applications}.
\newblock Cambridge University Press, 2005.

\bibitem{penny_gap}
F.~R. Capital.
\newblock The penny gap.
\newblock March 10, 2007.

\bibitem{dewan:HICSS}
R.~Dewan, M.~Freimer, and J.~Zhang.
\newblock Managing web sites for profitability: Balancing content and
  advertising.
\newblock {\em Hawaii International Conference on System Sciences}, 7, 2002.

\bibitem{falkinger:attention}
J.~Falkinger.
\newblock Attention economies.
\newblock {\em Journal of Economic Theory}, 133:266--294, 2007.

\bibitem{frederick}
S.~Frederick and G.~Loewenstein.
\newblock Hedonic adaptation.
\newblock In D.~Kahneman and E.~Diener, editors, {\em Well-being: The
  foundations of hedonic psychology}, pages 302--329. 1999.

\bibitem{frey:hapiness}
B.~S. Frey and A.~Stutzer.
\newblock What can economists learn from happiness research?
\newblock {\em Journal of Economic Literature}, 40(2):402--435, 2002.

\bibitem{godes:MS}
D.~Godes, E.~Ofek, and M.~Sarvary.
\newblock Content vs. advertising: The impact of competition on media firm
  strategy.
\newblock {\em Marketing Science}, 28(1):20--35, 2009.

\bibitem{wu}
B.~A. Huberman and F.~Wu.
\newblock Comparative advantage and efficient advertising in the attention
  economy.
\newblock {\em Proceedings of the 22nd European Conference on Operational
  Research (Euro XXII)}, 2007.

\bibitem{huberman:attention}
B.~A. Huberman and F.~Wu.
\newblock The economics of attention: maximizing user value in information-rich
  environments.
\newblock In {\em ADKDD '07: Proceedings of the 1st international workshop on
  Data mining and audience intelligence for advertising}, pages 16--20, New
  York, NY, USA, 2007.

\bibitem{kahneman:comp}
D.~Kahneman and R.~Thaler.
\newblock Economic analysis and the psychology of utility: Applications to
  compensation policy.
\newblock {\em The American Economic Review}, 81(2):341--346, May 1991.

\bibitem{anomalies}
D.~Kahneman and R.~H. Thaler.
\newblock Anomalies: Utility maximization and experienced utility.
\newblock {\em Journal of Economic Perspectives}, 20(1):221--�234, 2006.

\bibitem{kumar:control}
S.~Kumar and S.~P. Sethi.
\newblock Dynamic pricing and advertising for web content providers.
\newblock {\em European Journal of Operational Research}, 197:924–--944, 2009.

\bibitem{NYT:content}
J.~Lanier.
\newblock Pay me for my content.
\newblock {\em The New York Times}, November 20, 2007.

\bibitem{prasad:advertising}
A.~Prasad, V.~Mahajan, and B.~Bronnenberg.
\newblock Advertising versus pay-per-view in electronic media.
\newblock {\em Intern. J. of Research in Marketing}, 20:13–--30, 2003.

\bibitem{ariely:zero}
K.~Shampanier, N.~Mazar, and D.~Ariely.
\newblock Zero as a special price: The true value of free products.
\newblock {\em Marketing Science}, 26(6):742–--757, 2007.

\end{thebibliography}
